\documentclass[twocolumn]{aastex63}
\usepackage{url}
\usepackage{hyperref}

\received{July 23, 2018} 
\submitjournal{ApJ}

\shorttitle{Rapidly accreting black hole of PSO J006+39}
\shortauthors{Koptelova {\em et al.}}

\begin{document}

\defcitealias{2006ApJ...641..689V}{VP06}
\defcitealias{2017ApJ...839...93P}{P17}

\title{Rapidly accreting black hole of the Ly$\alpha$-luminous quasar PSO
J006.1240+39.2219}

\author{Ekaterina Koptelova}
\affiliation{Graduate Institute of Astronomy\\
National Central University \\
Taoyuan City 32001, Taiwan}


\author{Chorng-Yuan Hwang}
\affiliation{Graduate Institute of Astronomy\\
National Central University \\
Taoyuan City 32001, Taiwan}


\author{Matthew A.
Malkan}
\affiliation{Physics and Astronomy Department\\
University of California \\
Los Angeles, CA 90095-1547}


\author{Po-Chieh Yu}
\affiliation{Graduate Institute of Astronomy\\
National Central University \\
Taoyuan City 32001, Taiwan}


\begin{abstract}
We present near-infrared 1.1--1.3 and 1.3--1.6\,$\mu$m spectra of
the Ly$\alpha$-luminous quasar PSO J006.1240+39.2219 at
$z=6.617\pm0.003$ obtained with the NIRSPEC spectrograph at the
Keck-II telescope. The spectra cover the CIV $\it \lambda$1549,
CIII] $\lambda$1909 emission lines and part of the UV continuum of
the quasar. From the NIRSPEC observations of PSO
J006.1240+39.2219, we constrain the spectral slope of its UV
continuum to be $\alpha_{\rm \lambda}=-1.35\pm0.26$ and measure an
absolute magnitude of $M_{\rm 1450}=-25.60$. Using the scaling
relation between black hole mass, width of the CIV line and
ultraviolet continuum luminosity, we derive a black hole mass of
($2.19\pm0.30$)$\times$10$^{8}$$M_{\bigodot}$, which is consistent
but somewhat smaller than the typical black hole masses of
$z\gtrsim6$ quasars of similar luminosities. The inferred
accretion rate of $L_{\rm Bol}/L_{\rm Edd}\gtrsim2$ indicates that
PSO J006.1240+39.2219 is in the phase of the rapid growth of its
supermassive black hole characterized by the high NV/CIV line
ratio, $\rm NV/CIV>1$, and lower level of ionization of its
circumnuclear gas than in other high-redshift luminous quasars.
The NV/CIV line ratio of PSO J006.1240+39.2219 implies relatively
high abundance of nitrogen in its circumnuclear gas. This
abundance might be produced by the post-starburst population of
stars that provide the fuel for black hole accretion.
\end{abstract}

\keywords{quasars: emission lines --- quasars: supermassive black
holes --- quasars: individual (PSO J006.1240+39.2219)}

\section{Introduction}
\label{introduction}

The high-redshift quasars discovered at $z\gtrsim6$ are among the
most luminous known objects with typical absolute magnitudes of
$M_{\rm 1450}\thickapprox-26$
\citep{2016ApJS..227...11B,2016ApJ...833..222J}. There is also a
growing number of less luminous high-redshift quasars with $M_{\rm
1450}\gtrsim-25$
\citep[e.g.,][]{2010AJ....140..546W,2018ApJS..237....5M,koptelova2019}.
The generally high luminosities of high-redshift quasars are
powered by accretion onto massive supermassive black holes
(SMBHs). The masses of the SMBHs of these quasars are usually
estimated from their single-epoch spectra using the widths of the
CIV $\lambda$1549 and MgII $\lambda$2800 emission lines. The
relations between black hole mass and CIV and MgII line widths
have been calibrated for a sample of low-redshift active galactic
nuclei (AGNs) with their black hole masses directly measured from
reverberation mapping
\citep{2006ApJ...641..689V,2009ApJ...699..800V}. Assuming that
these relations can be extrapolated from local AGNs to the most
luminous distant quasars \citep[see,
e.g.,][]{2017ApJ...839...93P}, the derived black hole masses of
high-redshift quasars are typically on the order of
10$^{9}$$M_{\bigodot}$
\citep[e.g.,][]{2006AJ....132.2127J,2007AJ....134.1150J,2007ApJ...669...32K,2009ApJ...702..833K,2011ApJ...739...56D,2014ApJ...790..145D,2010AJ....140..546W,
2015Natur.518..512W,2019arXiv190407278O}. Recently,
\citet{2018Natur.553..473B} reported the discovery of the most
distant luminous quasar at $z=7.5$ (just 690\,Myr after the Big
Bang) with an SMBH of 8$\times$10$^{8}$$M_{\bigodot}$. The
existence of these SMBHs provides evidence of their extremely
rapid and efficient mass growth at early epochs. The discussed
scenarios of the mass growth of SMBHs are multiple mergers of
smaller initial black holes and gas accretion
\citep[e.g.,][]{2005ApJ...633..624V,2010A&ARv..18..279V,2014ApJ...784L..38M,2016MNRAS.456.2993L,2016MNRAS.458.3047P}.

The accretion rates of high-redshift quasars are typically between
0.1 and 1 of their Eddington luminosities
\citep{2011ApJ...739...56D,2014ApJ...790..145D,2015Natur.518..512W,2017ApJ...849...91M,2019arXiv190407278O}.
Some $z\gtrsim6$ quasars with less massive SMBHs
($\sim$10$^{8}$$M_{\bigodot}$) but accreting at the Eddington and
super-Eddington luminosities have also been discovered
\citep{2010AJ....140..546W,2017ApJ...849...91M}. They might be at
the early stages of the growth of their SMBHs. The observations of
these less massive, less evolved SMBHs can help us to test
different scenarios of the formation of SMBHs at high redshift.

The high-redshift quasars, including the most distant of them at
$z>7$ \citep{2011Natur.474..616M,2018Natur.553..473B}, exhibit the
same metal lines as quasars at low redshifts. The flux ratios of
different metal lines have dependency on the metallicity, density
and ionization state of the circumnuclear gas, and also on the
spectral energy distribution of the ionizing flux of quasars
\citep[e.g.,][]{1999ARA&A..37..487H,2006A&A...447..157N}. Assuming
that the gas properties of high-redshift quasars (density and
ionization state) are similar to those of low-redshift quasars,
the metal line ratios of high-redshift quasars indicate that their
gas metallicities are supersolar, possibly produced by
recent/ongoing star formation in their host galaxies
\citep[e.g.,][]{2007AJ....134.1150J,2009A&A...494L..25J,2014ApJ...790..145D}.
Indeed, the submillimeter observations of the hosts of
high-redshift quasars often show signs of a vigorous starburst
\citep[e.g.,][]{2009Natur.457..699W,2013ApJ...773...44W,2017ApJ...850..108W}.
However, the line flux ratios of high-redshift quasars may also
indicate that the gas density and ionization state of their
circumnuclear gas differ from those of low-redshift quasars.

The quasar PSO J006.1240+39.2219 at $z\thickapprox6.6$ (hereafter
PSO J006+39) selected by E.K. from the Panoramic Survey Telescope
and Rapid Response System 1 survey (PS1)
\citep{2002SPIE.4836..154K,2010SPIE.7733E..0EK,
2016arXiv161205560C} and confirmed with Subaru/FOCAS (Program
S15B0094N; PI: E. Koptelova) has unusual properties. The
Ly$\alpha$ line of PSO J006+39 is 3 to 4 times narrower
(FWHM$\sim$1300\,km\,s$^{-1}$) than a typical broad emission line
\citep[see][]{1997iagn.book.....P} and implies a less massive SMBH
than typically in the quasars known at $z>6.5$. At the same time,
the luminosity of PSO J006+39 of $M_{\rm 1450}\lesssim-26$,
inferred from its discovery spectrum,  is comparable to the
luminosities of other high-redshift luminous quasars at $z>6.5$
\citep{2017NatSR...741617K}. The Ly$\alpha$ $\lambda$1216 emission
line of PSO J006+39 is unusually strong relative to the continuum
\citep[see also][]{2017MNRAS.466.4568T}. We call PSO J006+39 a
Ly$\alpha$-luminous quasar as the luminosity of its Ly$\alpha$
line ($\sim$0.8$\times$10$^{12}$$L$$_{\bigodot}$) constitutes
almost 3\% of the total luminosity of PSO J006+39 and is larger
than the typical Ly$\alpha$ line luminosities of quasars by 2 to 3
times. We also found evidence of a significant quasar contribution
to the Ly$\alpha$ emission by observing fast variability of the
Ly$\alpha$ line of PSO J006+39 on timescales of days and weeks in
the quasar rest frame \citep{2017NatSR...741617K}. The narrow and
variable Ly$\alpha$ emission line of PSO J006+39 makes it similar
to local Narrow-line Seyfert 1 galaxies
\citep[NLSy1;][]{1985ApJ...297..166O}. NLSy1s exhibit narrow broad
emission lines as the result of their smaller SMBHs and often show
variability of UV lines and continuum
\citep[e.g.,][]{2001ApJ...561..146C,2002ApJ...564..162R}. Based on
the higher accretion rates and metallicities of NLSy1s in
comparison with broad line quasars
\citep[e.g.,][]{2000ApJ...542..692K,2002ApJ...567L..19S,2002ApJ...575..721N},
\citet{2000MNRAS.314L..17M} concluded that they might be at early
evolutionary stages similar to high-redshift quasars. The signs of
a recent star-formation activity in the host galaxies of
individual NLSy1s support the idea that these objects are
relatively young
\citep[e.g.,][]{2004ChJAA...4..415W,2006ApJ...648..158W}. The
properties of PSO J006+39 in comparison with local NLSy1s and
known high-redshift quasars can constrain the evolutionary stage
of its SMBH.

Here, we present new near-infrared observations of PSO J006+39
obtained at the Keck Observatory which cover the CIV
$\lambda$1549, CIII] $\lambda$1909 emission lines and part of the
UV continuum. Based on these observations, we report the first
measurement of the black hole mass and accretion rate of PSO
J006+39\footnote{During review of our paper,
\citet{2019MNRAS.484.2575T} published the results of their
analysis of the near-infrared spectrum of PSO J006+39 obtained
using Gemini North Near Infra-Red Spectrograph (GNIRS). As
suggested by our referee, we added a comparison between our
results derived from the Subaru/FOCAS and Keck/NIRSPEC data, and
from the Gemini/GNIRS data. The continuum and line properties of
PSO J006+39 estimated by us from the Gemini/GNIRS spectrum are
presented in Appendix~A.}. We also present the analysis of the
metallicity of the circumnuclear gas of PSO J006+39 using the flux
ratios of the observed emission lines. The results presented in
our paper provide new evidence that PSO J006+39 is in an early
phase of the black hole growth. In Section \ref{sec:observations},
we describe the observations and data reduction. In Section
\ref{sec:results}, we present analysis of the continuum and
emission lines of PSO J006+39, and derive the mass of its SMBH. In
Section \ref{sec:discussion}, we compare the black hole mass and
accretion rate of PSO J006+39 with those of other high-redshift
quasars and NLSy1s, and discuss its metal abundance. In Section
\ref{sec:conclusions} we present our main conclusions. In the
paper, we adopt the following cosmological parameters:
$H_{0}=67.8$\,km\,s$^{-1}$\,Mpc$^{-1}$, $\Omega_{M}=0.31$ and
$\Omega_{\Lambda}=0.69$ \citep{2016A&A...594A..13P}.

\begin{figure}[h!]
\vspace*{-12cm} \hspace*{-3.5cm} \centering
\includegraphics[scale=3.2]{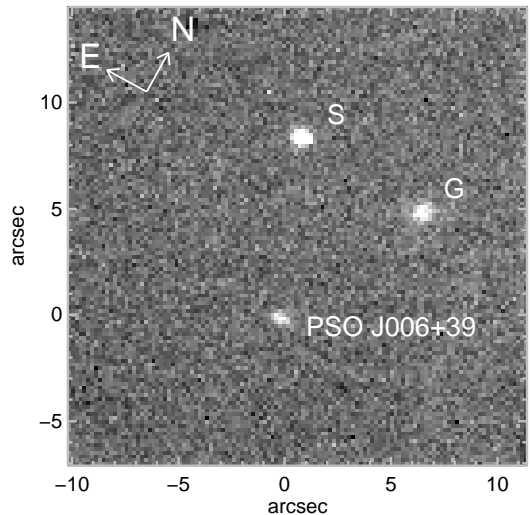}
\caption{NIRSPEC-6 $21\farcs6\times21\farcs$6 field of view of PSO
J006+39 with nearby star S and galaxy G marked.} \label{fig:fig1}
\end{figure}

\begin{figure*}[ht]
\centering
\includegraphics[width=\textwidth, clip, scale=1.5]{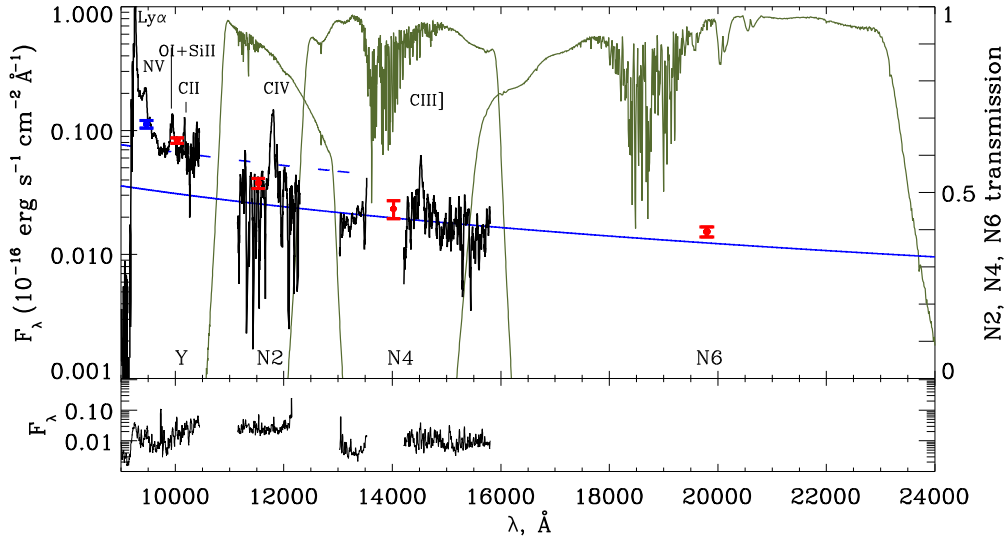}
\caption{Subaru/FOCAS and Keck/NIRSPEC spectrum of PSO J006+39
with the prominent emission lines marked. The displayed spectrum
is smoothed with a 10-pixel boxcar filter. Red points show the
fluxes of PSO J006+39 based on the FOCAS $Y$-band and NIRSPEC
$N2$-, $N4$-, $N6$-band photometry. The $y_{\rm PS1}$-band
multi-epoch flux of PSO J006+39 is shown by a blue point. The
transmission curves of the NIRSPEC $N2$, $N4$ and $N6$ filters are
overplotted. The power-law continuum fitted to the NIRSPEC
spectrum is shown with a blue line. The power-law continuum of the
FOCAS spectrum is marked with a dashed line. The 1-sigma error
spectrum in units of
10$^{-16}$\,erg\,s$^{-1}$\,cm$^{-2}$\,\AA$^{-1}$ is presented in
the lower panel. The FOCAS--NIRSPEC spectrum of this figure is
available as the Data behind the Figure.} \label{fig:fig2}
\end{figure*}

\section{OBSERVATIONS AND DATA REDUCTION}
\label{sec:observations}

The observations of PSO J006+39 were conducted using NIRSPEC, a
cross-dispersed, cryogenic, echelle spectrograph on the Keck-II
telescope \citep{1998SPIE.3354..566M}. The data were acquired on
the second half night of 2017 July 28 (UT) under
$\sim$$0\farcs55$--$0\farcs77$ seeing and cloudy sky with
decreasing cloud coverage during the observations. We used Low
Resolution mode of NIRSPEC with a 0$\farcs$76-wide and
42$\arcsec$-long slit which provided a resolving power of
$R\approx1100$ \citep[comparable to that of our previous
observations of PSO J006+39 with SUBARU/FOCAS,
see][]{2017NatSR...741617K}. The spectra of PSO J006+39 were taken
in the two bands, NIRSPEC-2 and NIRSPEC-4\footnote{The data are
available at \url{https://koa.ipac.caltech.edu/}}. The observed
wavelength intervals corresponding to our spectral setup were
1.1--1.3 and 1.3--1.6\,$\mu$m and included the CIV and CIII]
emission lines in the NIRSPEC-2 and NIRSPEC-4 bands, respectively.
During these observations, we took 8$\times$300\,s and
6$\times$300\,s exposures in NIRSPEC-2 and NIRSPEC-4,
respectively. The spectra of the quasar were taken simultaneously
with the spectra of a nearby star within $\sim$8$\farcs$6
north-east of the quasar by rotating the slit by an angle of
21$\fdg$6. This star is denoted as star S in
Figure~\ref{fig:fig1}. The PS1 mean $y$-band magnitude of star S
is $y_{\rm PS1}=19.65\pm0.07$\,AB\,mag. The $y_{\rm PS1}$-band
multi-epoch mean magnitude of PSO J006+39 which represents the
typical magnitude of the quasar measured over several PS1 epochs
is $y_{\rm PS1}=20.05\pm0.08$\,AB\,mag
\citep[see][]{2017NatSR...741617K}. Additionally, we took two 60-s
images of the quasar in each of the NIRSPEC-2, NIRSPEC-4 and
NIRSPEC-6 filters using a slit-viewing camera (SCAM). The SCAM
images had a pixel scale of 0\farcs18\,pixel$^{-1}$ and
21$\farcs$6-square field of view. The SCAM image of the quasar
field in the NIRSPEC-6 band is shown in Figure~\ref{fig:fig1}.

The spectra were reduced using IRAF\footnote{IRAF is distributed
by the National Optical Astronomy Observatories, which are
operated by the Association of Universities for Research in
Astronomy, Inc., under cooperative agreement with the National
Science Foundation.} tasks and IRAF-based package
WMKONSPEC\footnote{Available at \\
\url{https://www2.keck.hawaii.edu/inst/nirspec/wmkonspec.html}}.
The wavelength calibration was performed using night sky emission
lines from the quasar's frames. For the identification of the sky
lines we used the spectral atlas of \citet{2000A&A...354.1134R}.
The resulting dispersions of the NIRSPEC-2 and NIRSPEC-4 spectra
as measured from the sky lines were 2.179 and 2.894\,\AA\,
pixel$^{-1}$, respectively. The wavelength calibrated
2-dimensional frames of the quasar were sky corrected, aligned and
combined together. The 1D NIRSPEC-2 and NIRSPEC-4 spectra were
extracted from the combined 2-dimensional frames using IRAF task
$apall$. The continuum of PSO J006+39 was detected with low
signal-to-noise ratios of $SNR\approx1$ and 1.5--3.9 in the
NIRSPEC-2 and NIRSPEC-4 bands, respectively. Therefore, when
running $apall$, we used the spectrum of brighter star S as a
reference spectrum to trace the emission of PSO J006+39. The
resulting NIRSPEC-2 and NIRSPEC-4 spectra were corrected for
telluric lines and absolute flux calibrated using the telluric
standard stars, HIP8535 (type A1V, 2MASS $J=8.220\pm0.030$\,Vega
mag) and HIP114716 (type A0V, 2MASS $J=6.297\pm0.034$ Vega mag),
observed at similar airmass as the quasar. The accuracy of the
flux calibration of the resulting NIRSPEC-2 and NIRSPEC-4 spectra
is limited by the photometric accuracy ($\sim$0.03 magnitudes) of
the standard stars. Figure~\ref{fig:fig2} presents the spectrum of
PSO J006+39 from the Keck/NIRSPEC and previous Subaru/FOCAS
observations.

\tabcolsep=0.11cm
\begin{table}[ht]
\centering \caption{\label{table:t0} Photometry of PSO J006+39 in
the Subaru/FOCAS $Y$ band, and Keck/NIRSPEC $N2$, $N4$ and $N6$
bands.}
\begin{tabular}{lcccr@{$\pm$}l}
\hline \hline Band / Epoch & Exp. time& $\lambda_{\rm eff}$& Magnitude \\
                    & (s) &(\AA)& (AB mag) \\
\hline
\shortstack{FOCAS $Y$ / Nov 2, 2015}   &270     & 10036 &20.28$\pm$0.06            \\
\shortstack{NIRSPEC $N2$ / Jul 28, 2017}   & 120     & 11531 &20.84$\pm$0.10            \\
\shortstack{NIRSPEC $N4$ / Jul 28, 2017}   &120   & 14022 &20.94$\pm$0.18            \\
\shortstack{NIRSPEC $N6$ / Jul 28, 2017}   &120   & 19807 &20.65$\pm$0.09            \\
\hline
\end{tabular}
\end{table}

The photometric calibration of the quasar imaging data was
performed using near-infrared standard star Feige 22
\citep{2001MNRAS.325..563H} observed on the same night. The
NIRSPEC $N2$-, $N4$-, $N6$-band magnitudes of PSO J006+39 measured
from the SCAM images are summarized in Table~\ref{table:t0}. The
images of PSO J006+39 were also obtained in the FOCAS $Y$ band on
November 2, 2015 simultaneously with the spectrum of the quasar.
Table~\ref{table:t0} presents the FOCAS $Y$ band magnitude of PSO
J006+39 calibrated relative to the flux of BD+28d4211 observed
together with PSO J006+39 at similar airmass.

\section{RESULTS}
\label{sec:results}

\subsection{CONTINUUM}
\label{sec:continuum}

\begin{figure}[!ht]

\centering
\includegraphics[scale=1.8]{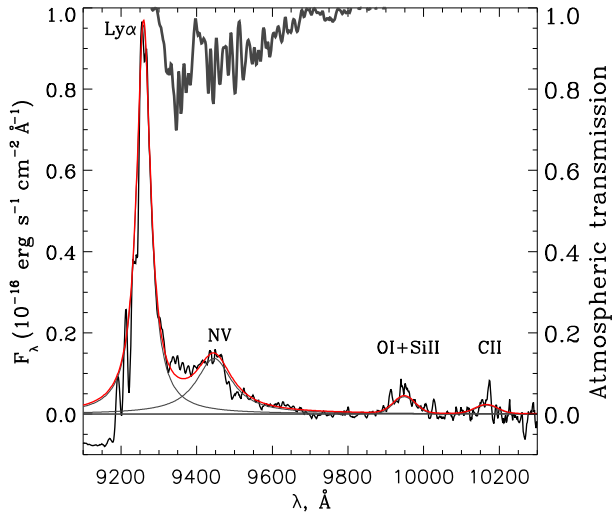}
\caption{Profiles of the Ly$\alpha$, NV, OI+SiII and CII emission
lines of PSO J006+39. The Ly$\alpha$ and NV lines were fitted with
Lorentzian profiles, and the OI+SiII and CII lines -- with
Gaussians after subtracting the power-law continuum. The fitted
Ly$\alpha$ and NV profiles are shown with grey lines, while the
total fit is shown by a red line. The atmospheric transmission
between 9300--9800\,\AA~ is overplotted with a solid thick line.}
\label{fig:fig3}
\end{figure}

In \cite{2017NatSR...741617K} we derived a spectral slope of the
PSO J006+39 UV continuum of $\alpha_{\lambda}=-1.10\pm0.48$ using
wavelength intervals 9500--9900 and 10000--10150\,\AA~ (where
$\alpha_{\lambda}$ is defined such that
$F_{\lambda}\propto\lambda^{\alpha_{\lambda}}$). In this previous
analysis, the FOCAS spectrum of PSO J006+39 was not corrected for
telluric absorption at wavelengths 9300--9800\,\AA.~ Moreover, the
9500--9900\,\AA~ wavelength interval includes the red wing of the
NV line which likely affected the previous slope measurement.
Here, we analyze the FOCAS spectrum corrected for telluric
absorption. The fraction of the quasar flux absorbed in the
earth's atmosphere was estimated by dividing the FOCAS spectrum of
the standard star BD+28d4211 by the flux-scaled black body model
of the star with an effective temperature of 82000\,K
\citep{2015A&A...579A..39L}. The derived atmospheric transmission
is displayed in Figure~\ref{fig:fig3}. We note that the $y_{\rm
PS1}$-band magnitude of PSO J006+39 estimated from the corrected
FOCAS spectrum is brighter by $\sim$0.1--0.3 than that estimated
by us previously in \citet{2017NatSR...741617K}.

The FOCAS and NIRSPEC spectra of PSO J006+39 were obtained at two
different epochs separated by 1 year and nine months (by slightly
less than 3 months in the quasar rest frame). Previously, we found
that the PS1 $y$-band light curve of PSO J006+39 shows brightness
variations with a peak-to-peak amplitude of $\sim$0.7 magnitudes
over $\sim$4 years \citep{2017NatSR...741617K}, which might be due
to the flux variations of both continuum and Ly$\alpha$ line of
PSO J006+39. To infer the brightness state of PSO J006+39 at the
epochs of its FOCAS and NIRSPEC observations, we first calculated
the spectral slope of the quasar continuum from the NIRSPEC
spectrum with a wider wavelength coverage than that of the FOCAS
spectrum. Using wavelength intervals of 11100--11300,
11400--11600, 13085--13400 and 14700--15200\,\AA~ we measured a
spectral slope of $\alpha_{\rm \lambda}=-1.35\pm0.26$, where the
quoted uncertainty is the statistical error of the fit. The fitted
power law is shown in Figure~\ref{fig:fig2} with a solid line. The
estimated continuum slope is consistent but somewhat flatter than
the typical slope of luminous quasars
\citep{1993ApJ...415..517Z,2001AJ....122..549V,2016A&A...585A..87S}.
We then fitted the FOCAS data using the power law with a fixed
spectral slope of $\alpha_{\lambda}=-1.35$ and spectral windows of
9700--9850 and 10050--10100\,\AA. The spectral windows adopted for
the analysis of the FOCAS and NIRSPEC spectra were taken to be
similar to the rest-frame wavelength intervals commonly used to
fit the continua of quasars
\citep{2001AJ....122..549V,2010MNRAS.402.2441D,2015MNRAS.449.4204L}
and less affected by the contribution from emission lines and the
Big Blue Bump (BBB) \citep[e.g.,][]{1983ApJ...268..582M}. The
estimated continuum flux of PSO J006+39 at the epoch of its FOCAS
observations is shown in Figure~\ref{fig:fig2} with a dashed line.
By comparing the continuum flux at the epochs of the FOCAS and
NIRSPEC observations, we find that the brightness state of PSO
J006+39 was different at these two epochs. PSO J006+39 was
brighter by about 0.8\,mag during the FOCAS observations than
during the NIRSPEC observations. Thus, the continuum flux of PSO
J006+39 might be different at different epochs depending on the
brightness state of the quasar. Figure~\ref{fig:fig2} also shows
the fluxes of PSO J006+39 in the FOCAS $Y$, and NIRSPEC $N2$, $N4$
and $N6$ bands at the epochs of the FOCAS and NIRSPEC observations
(see also Table~\ref{table:t0}).

\subsection{EMISSION LINES}

\begin{figure}[ht]
\centering \vspace{-105pt}
\includegraphics[width=\linewidth, clip, scale=2.6]{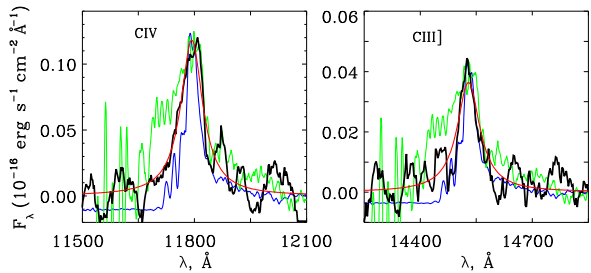}
\caption{CIV and CIII] emission lines of PSO J006+39 fitted with
Lorentzian profiles (red). The scaled profiles of the Ly$\alpha$
and NV lines are overplotted in blue and green for comparison. The
power-law continuum is subtracted from the line profiles. }
\label{fig:fig4}
\end{figure}

To estimate the fluxes and widths of the emission lines in the
FOCAS and NIRSPEC spectra of PSO J006+39, we subtracted the
power-law continuum ($F_{\lambda}\propto\lambda^{-1.35}$) and
fitted the lines with analytical profiles. The Ly$\alpha$, NV, CIV
and CIII] lines were fitted using Lorentzian profiles. The
intrinsic profile of Ly$\alpha$ is likely altered by neutral
hydrogen absorption seen as the series of absorption lines at its
blue side. However, these absorbtion features probably do not
significantly affect the total flux of the Ly$\alpha$ line due to
its intrinsically narrow width \citep[see][]{2017NatSR...741617K}.
The OI+SiII and CII emission lines were modelled using Gaussian
profiles. The fitted profiles of the emission lines are shown in
Figures~\ref{fig:fig3} and \ref{fig:fig4}. The estimated
properties of the lines are summarized in Tables~\ref{table:t1}
and \ref{table:t11}. From the line fit, we find that, similar to
the Ly$\alpha$ line, the CIV and CIII] emission lines of PSO
J006+39 are somewhat narrower compared with their usual widths in
broad line quasars ($>$2000\,km\,s$^{-1}$; see
Figure~\ref{fig:fig4}). The redshift of PSO J006+39 estimated as
the mean of the Ly$\alpha$+NV, OI+SiII, CIV and CIII] redshifts is
$z=6.617\pm0.003$. The estimated redshift is comparable within
uncertainties to that of the host galaxy of PSO J006+39 detected
in the [CII] 158\,$\mu$m emission by \citet{2017ApJ...849...91M}
($z_{\rm [CII]}=6.621\pm0.002$). However, the UV emission lines of
PSO J006+39 are seen slightly blueshifted by about
180--360\,km\,s$^{-1}$ relative to the [CII] 158-$\mu$m line.

The absolute magnitude of the continuum of PSO J006+39 at rest
frame wavelength 1450\,\AA~estimated from the NIRSPEC data is
$M_{1450}=-25.60\pm0.07$ magnitudes, where the error includes a
flux calibration uncertainty of 0.03~mag and an uncertainty of
0.06\,mag introduced by the slope error added in quadrature. The
uncertainty introduced by the slope error was estimated from a
sample of simulated NIRSPEC spectra with the continuum slopes
normally distributed around a mean of --1.35 and a standard error
of 0.26. We also note that the absolute magnitude of PSO J006+39
at the epoch of the FOCAS observations was $M_{1450}=-26.43$
(assuming a slope of $\alpha_{\rm \lambda}=-1.35$).

\tabcolsep=0.11cm
\begin{deluxetable}{lccr@{$\pm$}lr@{$\pm$}lr@{$\pm$}l}
\tabletypesize{\footnotesize}
\tablewidth{\textwidth} \tablecaption{Properties of the emission
lines. \label{table:t1}}

\tablehead{\colhead{Line}                          &
           \colhead{$\lambda_{\rm peak}$}                          &
           \colhead{Redshift}                      &
           \multicolumn{2}{c}{10$^{-16}\times$$F_{\rm line}$}                            &
           \multicolumn{2}{c}{EW}              &
           \multicolumn{2}{c}{FWHM}              \\
           \colhead{}                              &
           \colhead{(\AA)}                              &
           \colhead{}                      &
           \multicolumn{2}{c}{(erg s$^{-1}$cm$^{-2}$)}             &
           \multicolumn{2}{c}{(\AA)}             &
           \multicolumn{2}{c}{(km s$^{-1}$)}             }
\startdata
Ly$\alpha$      & $ 9259.7$ & $6.617$               & $ 67.1$ & $0.8$             &         $114$ & $ 1$ &      $1445$ & $ 30$  \\
NV              & $ 9446.1$ & $6.617$               & $ 24.7$ & $0.9$             &         $39$ & $ 2$ &      $3657$ & $ 85$ \\
OI+SiII         & $ 9948.9$ & $6.620$               & $ 3.1$  & $0.3$             &         $6$ & $ 2$ &      $2039$ & $ 169$ \\
CII             & $ 10165.9$ & $--$                 & $ 1.7$  & $0.5$             &         $3$ & $ 3$ &       $2085$ & $ 336$ \\
CIV             & $ 11792.9$ & $6.613$              & $ 13.2$  & $0.6$             &         $57$ & $ 3$ &      $1810$ & $ 119$ \\
CIII]           & $ 14530.4$ & $6.613$              & $ 4.2$  & $0.8$             &         $20$ & $ 5$ &      $1515$ & $ 186$ \\
\enddata
\end{deluxetable}

\begin{deluxetable}{cr@{$\pm$}lr@{$\pm$}lr@{$\pm$}lr@{$\pm$}l}
\tabletypesize{\footnotesize}
\tablewidth{\textwidth} \tablecaption{Fluxes of the Ly$\alpha$, NV
and CIII] lines relative to the flux of the CIV line (see
Table~\ref{table:t1}) measured from the FOCAS and NIRSPEC spectra
of PSO J006+39. \label{table:t11}}

\tablehead{\colhead{Line}                          &
           \multicolumn{2}{c}{Ly$\alpha$} &
           \multicolumn{2}{c}{NV}  &
           \multicolumn{2}{c}{CIII]}}
\startdata $F$$_{\rm line}$/$F$$_{\rm CIV}$ & $ 5.1$ &
$0.7$\tablenotemark{a}   & $ 1.9$ & $0.3$\tablenotemark{a} & $
0.3$ & $0.1$
\enddata
\tablenotetext{a}{The uncertainties take into account possible
changes in the line fluxes by up to 10\% between the epochs of the
Subaru/FOCAS and Keck/NIRSPEC observations. }
\end{deluxetable}

\subsection{BLACK HOLE MASS}

\begin{table}[ht]
\centering \caption{\label{table:t2} $\alpha_{\lambda}$, $M_{\rm
1450}$, $\lambda$$L_{\rm \lambda}$(1350\AA), $\lambda$$L_{\rm
\lambda}$(3000\AA), $M_{\rm BH}$ and $L_{\rm Bol}/L_{\rm Edd}$
derived from the Keck/NIRSPEC spectrum of PSO J006+39.}
\begin{tabular}{lc}
\hline \hline
$\alpha_{\rm \lambda}$              & --1.35$\pm$0.26            \\
$M$$_{\rm 1450}$              & --25.60$\pm$0.07            \\
$\lambda$ $L_{\rm \lambda}$(1350\AA)  (10$^{46}$ erg s$^{-1}$)           &       1.57$\pm$0.12    \\
$\lambda$ $L_{\rm \lambda}$(3000\AA)  (10$^{46}$ erg s$^{-1}$)           &       1.18$\pm$0.15    \\

$M_{\rm BH}$ (10$^{8}$ $M_{\bigodot}$), \citetalias{2006ApJ...641..689V} & 2.19$\pm$0.30           \\
$L_{\rm Bol, 1350}/L_{\rm Edd}$, \citetalias{2006ApJ...641..689V}           & 2.17$\pm$0.34           \\
$L_{\rm Bol, 3000}/L_{\rm Edd}$, \citetalias{2006ApJ...641..689V}           & 2.21$\pm$0.41           \\

$M_{\rm BH}$ (10$^{8}$ $M_{\bigodot}$), \citetalias{2017ApJ...839...93P}  & 1.20$\pm$0.16           \\
$L_{\rm Bol, 1350}/L_{\rm Edd}$, \citetalias{2017ApJ...839...93P}           & 3.95$\pm$0.60           \\
$L_{\rm Bol, 3000}/L_{\rm Edd}$, \citetalias{2017ApJ...839...93P}           & 4.02$\pm$0.74           \\

\hline
\end{tabular}
\end{table}

To estimate the mass of the SMBH of PSO J006+39, we used the
empirical relation between black hole mass, CIV line width and UV
continuum luminosity found by \citet[][hereafter
VP06]{2006ApJ...641..689V}. This relation has been used for the
estimation of the black hole masses of other high-redshift quasars
which allows for the direct comparison of our results with the
previous works. The intrinsic scatter of this relation is 0.36
dex. It is based on the virial equation in which black hole mass
is proportional to the size of the broad line region ($R$) and to
the square of the emission line width (FWHM$^{2}$). The size of
the broad line region of high-redshift quasars is generally
unknown. It is estimated using the empirical relation between $R$
and the continuum luminosity of the form $R \varpropto L^{\beta}$
found from reverberation mapping of local AGNs ($\beta$ is equal
to 0.5 in the relation of \citetalias{2006ApJ...641..689V}).

The method of the black hole mass measurement from broad emission
lines of quasars relies on the assumption that the broad line
region is virialized. However, the profile of the CIV line of
quasars is often altered by non-virial effects (outflows or
absorption in the circumnuclear region) which make the CIV line a
less reliable black hole mass estimator compared to low-ionization
emission lines such as MgII. As discussed in
\citetalias{2006ApJ...641..689V} and \cite{2017MNRAS.465.2120C},
the blueshifted by $>$1000\,km\,s$^{-1}$ CIV line generally leads
to overestimation of the CIV black hole masses by a few times.

The CIV line of PSO006+39 is symmetric and is not significantly
blueshifted with respect to the other UV emission lines. This
suggests that any possible bias in the black hole mass measurement
due to the blueshift of the CIV line is likely small. The
estimated black hole mass and Eddington ratio of PSO J006+39 are
summarized in Table~\ref{table:t2}. The quoted errors include the
uncertainties in the line width of the CIV line and monochromatic
luminosity $\lambda$$L_{\rm \lambda}$(1350\AA). We estimated the
bolometric luminosity of PSO J006+39 by applying bolometric
correction factors of 3.81 and 5.15 to the continuum luminosities
at 1350 and 3000~\AA,~respectively
\citep{2006ApJS..166..470R,2008ApJ...680..169S}. We note that the
estimated continuum luminosities correspond to the brightness
state of PSO J006+39 at the epoch of the Keck/NIRSPEC observations
and might change with time (see Section~\ref{sec:results}). The
Eddington luminosity was derived as $L_{\rm Edd}=1.26\times
10^{38}$($M_{\rm BH}$/$M_{\bigodot}$)\,erg\,s$^{-1}$
\citep[e.g.,][]{1997iagn.book.....P}. The accretion rate of PSO
J006+39 presented in Table~\ref{table:t2} was estimated for two
values of the bolometric luminosity, $L_{\rm Bol, 1350}$ and
$L_{\rm Bol, 3000}$, derived from the continuum luminosity at 1350
and 3000\,\AA, respectively. In previous works, the bolometric
luminosities of high-redshift quasars were estimated from their
continuum luminosities at 3000\,\AA~ by applying a bolometric
correction factor of 5.15
\citep[see][]{2011ApJ...739...56D,2014ApJ...790..145D,2017ApJ...849...91M}.
In Section~\ref{sec:discussion}, to compare the accretion rate of
PSO J006+39 with those of other high-redshift quasars, we use the
bolometric luminosity estimated in the same way.
\citet{2014ApJ...790..145D} used both CIV and MgII emission lines
and relations of \citetalias{2006ApJ...641..689V} and
\citet{2009ApJ...699..800V} to measure the black hole masses of a
few quasars at $z\gtrsim6.5$. The CIV and MgII black hole masses
of these quasars are consistent within 0.10--0.34\,dex. We note
that unlike the CIV line of PSO J006+39, the CIV lines of these
quasars are significantly blueshifted by
$\gtrsim$1000\,km\,s$^{-1}$ \citep[see
also][]{2017ApJ...849...91M}. Figure 5 presents the comparison of
the CIV black hole masses and accretion rates of these quasars and
those of PSO J006+39.

We also derived the black hole mass of PSO J006+39 using the
relation of \citet[][hereafter P17]{2017ApJ...839...93P} found for
a sample of low--redshift AGNs with high-quality HST spectra (see
Table~\ref{table:t2}). \citetalias{2017ApJ...839...93P} considered
two cases, when black hole mass depends on line width as
FWHM$^{2}$ and when it depends as FWHM$^{\gamma}$, allowing
$\gamma$ to be different from a physically expected value of 2.
For our comparison, we used the \citetalias{2017ApJ...839...93P}
relation found for $\gamma$ fixed to 2 which has an intrinsic
scatter of 0.43 dex. The black hole mass of PSO J006+39 derived
from the relations of \citetalias{2006ApJ...641..689V} and
\citetalias{2017ApJ...839...93P} differs by a factor of two which
reflects the typical systematic uncertainties in black hole mass
measurements using different relations and emission lines. The
uncertainties in the black hole mass and accretion rate of PSO
J006+39 related to continuum variability are also within the
typical systematic uncertainties.

\section{DISCUSSION}
\label{sec:discussion}
\subsection{Accretion rate}
\label{sec:rate}

\begin{figure}[!ht]
\centering \vspace{-120pt}
\includegraphics[scale=0.68]{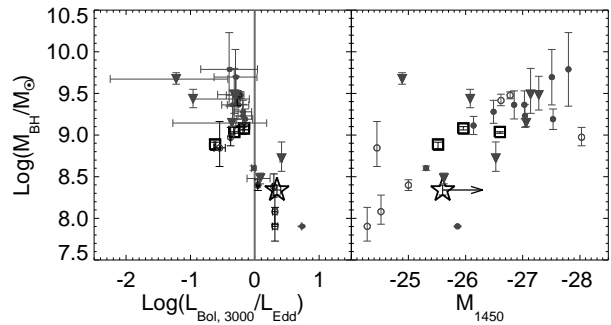}
\caption{Black hole mass as a function of the Eddington fraction
and absolute magnitude. PSO J006+39 is shown with an open star.
The arrow indicates the high luminosity brightness states of PSO
J006+39. The $z>6.5$ quasars from \cite{2014ApJ...790..145D} with
black hole masses derived from the CIV line are shown with open
squares. The quasars from \citet{2010AJ....140..546W},
\citet{2011ApJ...739...56D} and \cite{2017ApJ...849...91M} with
the black hole masses derived from the MgII line are shown with
open and filled circles, and triangles for comparison. The solid
vertical line corresponds to an Eddington fraction of 1.}
\label{fig:fig5}
\end{figure}

The early analysis of the UV-to-infrared spectra of quasars and
Seyfert galaxies by \citet{1989ApJ...346...68S} showed that
luminous quasars typically have higher accretion rates than
low-redshift Seyfert galaxies. The increasing number of quasars
found in different surveys at $z>1$ revealed that their Eddington
ratios are typically between $-1<\rm log(\it L_{\rm Bol}/\it
L_{\rm Edd})<\rm0$
\citep{2009ApJ...699..800V,2008ApJ...680..169S}. In
Figure~\ref{fig:fig5}, we show the black hole masses versus
Eddington ratios and absolute magnitudes of high-redshift quasars
from the samples of
\citet{2010AJ....140..546W,2011ApJ...739...56D,2014ApJ...790..145D,
2017ApJ...849...91M}.  As seen in the figure, the accretion rates
of high-redshift quasars with SMBHs of
$\gtrsim10^{9}$$M$$_{\bigodot}$ are also typically between $-1<\rm
log$($L_{\rm Bol}/L_{\rm Edd})<0$. There are also less luminous
high-redshift quasars with luminosities of $M_{\rm 1450}>-26$ and
SMBHs of $<10^{9}$$M$$_{\bigodot}$ which exhibit higher accretion
rates \citep[e.g.,][]{2010AJ....140..546W}.
\citet{2011ApJ...739...56D} analyzed a sample of 19 quasars at
$4.5<z<6.4$ including nine low luminosity quasars from
\citet{2010AJ....140..546W}. For their sample,
\citet{2011ApJ...739...56D} found a higher mean Eddington ratio
than for luminosity-matched quasars at low redshifts, log$(L_{\rm
Bol}$/$L_{\rm Edd})=-0.37\pm0.20$ and log$(L_{\rm Bol}$/$L_{\rm
Edd})=-0.8\pm0.24$, respectively. The recent analysis of 11
$z\gtrsim6.5$ quasars by \citet{2017ApJ...849...91M} did not show
significant difference between Eddington ratios and black hole
masses of high-redshift and luminosity-matched low-redshift
quasars.

As seen in Figure~\ref{fig:fig5}, the luminosity of PSO J006+39 is
similar to the luminosities of the known luminous quasars at $z>
6.5$ ($M_{\rm 1450}\sim-26$). However, the mass of the SMBH of PSO
J006+39 is 3--4 times smaller than the SMBH masses of these
$z>6.5$ quasars. The high accretion rate of the SMBH of PSO
J006+39 compared to other luminous quasars might suggest that it
is in an unusual phase of the rapid growth
\citep{2016MNRAS.458.3047P,2016MNRAS.456.2993L}. The efficiency of
the growth of SMBHs is expected to decline at masses of $\sim$
10$^{9}$$M$$_{\bigodot}$ caused by gas depletion due to SMBH
accretion, quasar outflow and star formation
\citep[e.g.,][]{2002ApJ...564..592H}. The high accretion rate of
PSO J006+39 indicates that there is probably plenty of gas
surrounding its SMBH which provides material for its accretion.
The presence of the strong outflow that moves gas away from a SMBH
might be indicated by the blueshift of the CIV line. In the known
luminous high-redshift quasars, the CIV line is usually
significantly blueshifted by $\gtrsim$1000\,km\,s$^{-1}$
\citep[e.g.,][]{2014ApJ...790..145D,2017ApJ...849...91M}. Compared
to these quasars, the CIV line of PSO J006+39 does not show any
significant blueshift. Thus, the outflow of gas in PSO J006+39 is
probably not too strong to prevent its SMBH from the active
growth.

The accretion rate of PSO J006+39 and the amplitude of its
intrinsic brightness variations are consistent with the accretion
rate--variability distribution of low-luminosity quasars ($L_{\rm
Bol}<10^{46}$\,erg\,s$^{-1}$;
\citealp[e.g.,][]{2019ApJ...877...23L}). The amplitude of the
intrinsic brightness variations of PSO J006+39 estimated from its
$y_{\rm PS1}$-band light curve covering $\sim$5 months in the
quasar rest frame \citep{2017NatSR...741617K}, is $\sim$24\%
\citep[calculated as in][]{2002ApJ...568..610E}. This amplitude is
also consistent (although slightly higher) with brightness
variations of $\sim$15--19\% expected from the dependence of the
structure function of quasars on their physical parameters such as
luminosity and black hole mass
\citep[e.g.,][]{2010ApJ...721.1014M}.

\begin{figure}[ht]
\centering
\includegraphics[width=\linewidth, clip, scale=2.]{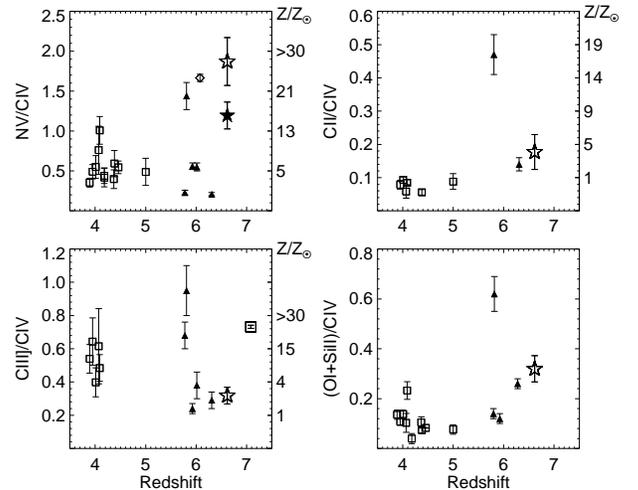}
\caption{NV/CIV, CII/CIV, CIII]/CIV, and (OI+SiII)/CIV line ratios
as a function of redshift. The right axes show the approximate
metallicities estimated using Table 10 of
\citet{2006A&A...447..157N}. The NV/CIV line ratio of PSO J006+39
estimated from the FOCAS--NIRSPEC and GNIRS data is marked with
open and filled stars, respectively. The sample of $3.9<z<5.0$
quasars from \citet{2003A&A...398..891D} is shown with squares.
The quasars at $z\sim6$ from \citet{2007AJ....134.1150J} are shown
with triangles. The NV/CIV line ratio of the quasar SDSS
J0303-0019 at $z=6.08$ from \citet{2009ApJ...702..833K} is marked
with a diamond. The CIII]/CIV line ratio of ULAS J1120+0641 at $z=
7.1$ from \citet{2014ApJ...790..145D} is marked with an open
square.} \label{fig:fig6}
\end{figure}

\subsection{Metal abundance and ionization of \\circumnuclear gas}

It is usually assumed that the density and ionization state of the
circumnuclear gas in low- and high-redshift quasars are similar
\citep{1999ARA&A..37..487H,2006A&A...447..157N}. If this
assumption is correct, the flux ratios between different UV lines
of high-redshift quasars indicate that their gas metallicities are
3--10 times of solar metallicity even at redshift $z>6$
\citep{2002AJ....123.2151P,2003A&A...398..891D,2003ApJ...589..722D,2007AJ....134.1150J,
2009ApJ...702..833K}. Moreover, the relative abundances of metals
seem to have no dependence on redshift
\citep{2002AJ....123.2151P,2007ApJ...669...32K,2011ApJ...739...56D,2014ApJ...790..145D,2009A&A...494L..25J,2018MNRAS.480..345X}.
In particular, no redshift evolution was found for the FeII/MgII
line ratio up to redshift $z=7.1$ \citep[see,
e.g.,][]{2014ApJ...790..145D}, suggesting that the metal
enrichment of high-redshift quasars happens rapidly and mostly
ends before their luminous phase \citep{2002ApJ...564..592H}.

To infer the metal abundance of PSO J006+39, we examined the
following flux ratios: NV/CIV, CIII]/CIV, CII/CIV, and
(OI+SiII)/CIV presented in Figure~\ref{fig:fig6}. The NV/CIV,
CII/CIV, and (OI+SiII)/CIV line ratios involve emission lines
observed with FOCAS and NIRSPEC at two different epochs. In
general, the fluxes of the same emission lines observed at
different epochs could be different. In particular, the
high-ionization lines such as NV and CIV might be more variable
compared to the CII and OI+SiII lines. The typical brightness
change of high-ionization lines in some strongly variable quasars
is about 10\% over a few years
\citep[e.g.,][]{2018ApJ...865...56L}. To account for the
variations of the NV and CIV line fluxes of PSO J006+39, we added
in quadrature 10\% of the measured line fluxes to the flux
uncertainties of these lines quoted in Table~\ref{table:t1}. Thus,
the errorbars shown in Figure~\ref{fig:fig6} include uncertainties
in the measured line fluxes and uncertainties due to the variation
of the NV and CIV line fluxes by up to 10\% between the epochs of
the FOCAS and NIRSPEC observations. Figure~\ref{fig:fig6} also
shows the flux ratio between the NV and CIV lines derived by us
from the Gemini/GNIRS data in Appendix~A. The fluxes of the NV and
CIV lines from the GNIRS spectrum were measured at the same epoch
and should not be affected by the flux variations of these two
lines. Therefore, the NV/CIV flux ratio from the GNIRS spectrum
likely represents the typical NV/CIV ratio of PSO J006+39. As seen
in Figure~\ref{fig:fig6}, both GNIRS and FOCAS/NIRSPEC estimates
indicate a relatively high NV/CIV ratio in PSO J006+39 than
typically in high-redshift quasars.

To estimate the metallicity of PSO J006+39 from the NV/CIV,
CIII]/CIV, and CII/CIV line ratios, we used the results of
calculations by \citet{2006A&A...447..157N}. Using the locally
optimally emitting cloud model of \citet{1995ApJ...455L.119B},
\citet{2006A&A...447..157N} calculated fluxes of different
emission lines relative to the flux of CIV produced by gas clouds
of different densities and at different radii, and for the gas
metallicity in a range of 0.2 and 10$Z$$_{\bigodot}$ \citep[see
Table 10 of][]{2006A&A...447..157N}. To derive the metallicity
using the observed line flux ratios, we extrapolated the model
predicted line ratios of \citet{2006A&A...447..157N}.
Figure~\ref{fig:fig6} shows the approximate metallicity of PSO
J006+39 corresponding to measured line ratios NV/CIV, CIII]/CIV,
and CII/CIV. For comparison, Figure~\ref{fig:fig6} also shows the
approximate metallicities of the known high-redshift quasars from
\citet{2003A&A...398..891D,2007AJ....134.1150J,2009ApJ...702..833K,2014ApJ...790..145D}
also estimated based on the calculations of
\citet{2006A&A...447..157N}. The relative abundance of nitrogen is
proportional to metallicity and therefore the line ratios such as
NV/CIV are often used as metallicity indicators of the broad line
regions of quasars \citep{1993ApJ...418...11H}. The NV/CIV line
ratio of PSO J006+39 of $>$1 implies a gas metallicity of
$>$10$Z_{\bigodot}$. The CIII]/CIV and CII/CIV line ratios
resulting in metallicities of $\sim$2$Z_{\bigodot}$ and
$\sim$4$Z$$_{\bigodot}$ might not be reliable metallicity
indicators since the CIII] and CII lines originate at different
emitting regions with the CIV line
\citep[e.g.,][]{2002ApJ...564..592H}. Besides, the CII/CIV and
also (OI+SiII)/CIV line ratios might be increasing with redshift
as seen for the sample of \citet{2007AJ....134.1150J}. Earlier,
\citet{2006A&A...447..157N} noted that the (OI+SiII)/CIV line
ratio might marginally correlate with redshift \citep[see Figure
24 of][]{2006A&A...447..157N}.

\begin{figure}[!ht]

\centering \vspace{-120pt}
\includegraphics[scale=0.68]{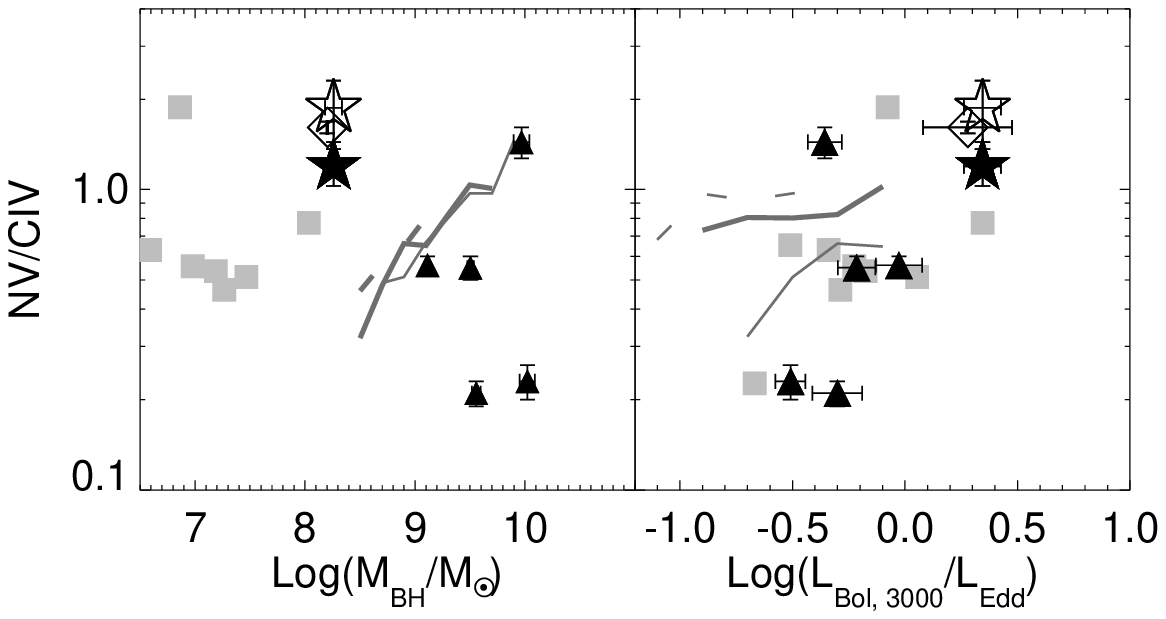}
\caption{NV/CIV line ratio vs. black hole mass and accretion rate.
The NV/CIV line ratio of PSO J006+39 estimated from the
FOCAS--NIRSPEC and GNIRS data is marked with open and filled
stars, respectively. The quasars at $z\sim6$ from
\citet{2007AJ....134.1150J} and from \citet{2009ApJ...702..833K}
are shown with triangles and a diamond, respectively. NLSy1
galaxies from \citet{2002ApJ...567L..19S, 2004ApJ...614..547S} are
shown with grey squares for comparison. The lines show the
observed relations between the NV/CIV line ratio, black hole mass
and accretion rate from \citet{2011A&A...527A.100M}. The dashed,
thick and thin lines in the left panel correspond to median
accretion rates log($L_{\rm Bol}/L_{\rm Edd})=-0.1$, $-0.3$,
$-0.5$. The dashed, thick and thin lines in the right panel
correspond to median black hole masses log($M_{\rm
BH}/M_{\bigodot})=9.7$, 9.3, 8.9.} \label{fig:fig7}
\end{figure}

The analysis of a large sample of quasars by
\citet{2011A&A...527A.100M} also showed that the NV/CIV line ratio
tends to be larger for quasars with more massive SMBHs or higher
accretion rates \citep[see
also][]{2003ApJ...596...72W,2004ApJ...614..547S,2017A&A...608A..90M,2018MNRAS.480..345X}.
The dependance of the NV/CIV line ratio on black hole mass may
possibly result from the mass-metallicity relation of galaxies
\citep[e.g.,][]{2004ApJ...613..898T}. However, its dependence on
accretion rate is less clear. \citet{2011A&A...527A.100M}
suggested that mass accretion rates onto growing SMBHs might be
associated with the post-starburst phase when the mass loss of the
post-starburst population of stars (AGB stars) triggers quasar
activity. These stars can quickly enrich the central regions of
quasars' host galaxies with nitrogen. Nitrogen might be then
ingested into the broad line regions of quasars with stellar winds
fuelling the black hole accretion \citep[see
also][]{2007ApJ...671.1388D}. Given this scenario, the high NV/CIV
line ratio of PSO J006+39 might be due to the local overabundance
of nitrogen rather than due to the overall high metallicity of its
broad line region and the central region of its host galaxy. The
quasars showing high NV/CIV line ratios are found to be rare at
any redshift which might suggest that this evolutionary phase is
relatively short \citep{2012A&A...543A.143A,2017A&A...608A..90M}.
NLSy1 galaxies also typically exhibit somewhat high NV/CIV line
ratios for their small black hole masses
\citep{2002ApJ...567L..19S,2004ApJ...614..547S}. In
Figure~\ref{fig:fig7}, we show the NV/CIV line ratio of PSO
J006+39 in comparison with that of the quasars at $2.3<z<3.0$ from
\citet{2011A&A...527A.100M}, high-redshift quasars at $z\sim6$ and
NLSy1s. As seen in this figure, the NV/CIV line ratio of PSO
J006+39 is somewhat high for its black hole mass, but follows the
trend of increasing NV/CIV line ratio with accretion rate. Using
different emission and absorption lines, the metallicities of the
interstellar media in the central regions of NLSy1s are usually
estimated to be 1--5$Z_{\bigodot}$
\citep{2002ApJ...567L..19S,2002ApJ...575..721N,2005ApJ...634..928F}.
Given the relatively small SMBH of PSO J006+39 compared to other
$z>6.5$ quasars and its similarity with NLSy1s, the overall
metallicity of the central region of its host galaxy is likely a
few times smaller than that indicated by the NV/CIV line ratio.

\begin{figure}[ht]
\centering
\includegraphics[width=\linewidth, clip, scale=2.8]{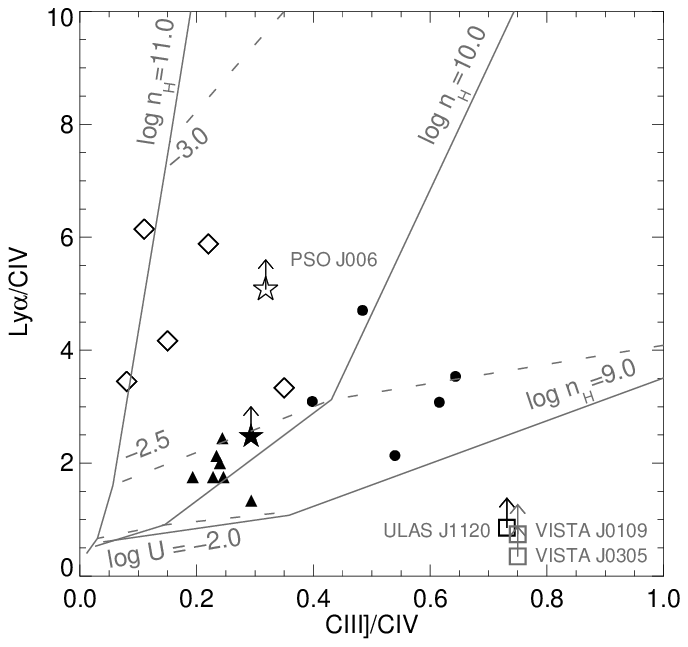}
\caption{CIII]/CIV vs. Ly$\alpha$/CIV. The line ratios of PSO
J006+39 from the FOCAS and NIRSPEC, and GNIRS data are shown with
open and filled stars, respectively. The line ratios of NLSy1
galaxies from \citet{2000ApJ...542..692K} (open diamonds), of
quasars at $3.9<z<5.0$ from \citet{2003A&A...398..891D} (solid
circles) and the average line ratios of quasars at $0<z<5.0$ from
\citet{2004ApJ...608..136W} (triangles) are shown for comparison.
The quasar ULAS J1120+0641 at $z=7.1$ is shown with an open square
\citep{2014ApJ...790..145D}. The two grey squares mark the quasars
VISTA J0109-3047 at $z=6.75$ and VISTA J0305-3150 at $z=6.61$
\citep[see][]{2014ApJ...790..145D}. (As their CIII]/CIV line
ratios are not measured, we assumed that they are close to those
of ULAS J1120+0641 and PSO J006+39). The measured Ly$\alpha$/CIV
line ratios of PSO J006+39, ULAS J1120+0641, VISTA J0109-3047 and
VISTA J0305-3150 represent lower limits, since the Ly$\alpha$ flux
of high-redshift quasars is typically reduced by neutral hydrogen
absorption. The grid of the theoretical line ratios calculated by
\citet{2000ApJ...542..692K} for gas densities log $n$$_{\rm
H}=9.0, 10.0, 11.0$ and ionization parameters log $U=-2.0, -2.5,
-3.0$ are shown with solid and dashed lines. } \label{fig:fig8}
\end{figure}

The ratios between the Ly$\alpha$, CIII] and CIV lines are
sensitive to the density and ionization state of the circumnuclear
gas. In Figure~\ref{fig:fig8}, we plot the Ly$\alpha$/CIV and
CIII]/CIV line ratios of PSO J006+39 in comparison with the line
ratios of $z<5$ quasars from
\citet{2003A&A...398..891D,2004ApJ...608..136W}, NLSy1 galaxies
from \citet{2000ApJ...542..692K} and high-redshift quasars at
$z>6.5$ from \citet{2014ApJ...790..145D}. The figure also shows
the model-predicted line ratios calculated for different values of
the gas density and ionization parameter by
\citet{2000ApJ...542..692K} to explain the line ratios of local
NLSy1s. As seen in this figure, the Ly$\alpha$/CIV and CIII]/CIV
line ratios of PSO J006+39 correspond to a lower ionization
parameter of the circumnuclear gas in comparison with the known
$z>6.5$ quasars from \citet{2014ApJ...790..145D}. Moreover, the
Ly$\alpha$/CIV line ratio from the FOCAS--NIRSPEC, and GNIRS data
indicates that the ionization parameter of PSO J006+39 may vary
significantly depending on the brightness state of the quasar. The
low level of gas ionization due to the high density of gas is more
typical for the broad line regions of NLSy1s. As also seen in
Figure~\ref{fig:fig8}, the gas density in the broad line region of
PSO J006+39 is consistent with the typical gas densities in the
broad line regions of quasars at $z<5$. Therefore, the overall
metal abundance of its circumnuclear gas should not be too
different from the typical metal abundances of quasars except for
the high nitrogen abundance. The density of the circumnuclear gas
of PSO J006+39 seems to be one order higher than typically in the
known quasars at $z>6.5$ (see Figure~\ref{fig:fig8}). We note that
the actual density/ionization state of the circumnuclear gas of
these $z>6.5$ quasars might be somewhat higher than that indicated
by their Ly$\alpha$/CIV line ratios as their Ly$\alpha$ line flux
is usually severely reduced by neutral hydrogen absorption. The
inferred density of gas and low level of ionization similar to
that of NLSy1s might imply that PSO J006+39 is at the early
evolutionary stage of luminous quasars.

\section{CONCLUSIONS}
\label{sec:conclusions}

We presented the new Keck/NIRSPEC observations and combined
analysis of the Subaru/FOCAS and Keck/NIRSPEC data of the
Ly$\alpha$-luminous quasar PSO J006.1240+39.2219 at
$z=6.617\pm0.003$. Based on these observations, we measured the
black hole mass and accretion rate of PSO J006+39. We analyzed the
metallicity and ionization state of the circumnuclear gas of PSO
J006+39 using the emission lines covered by the Subaru/FOCAS and
Keck/NIRSPEC spectra. We summarize our results as follows:

1. From the analysis of the Keck/NIRSPEC spectrum of PSO J006+39,
we measured the spectral slope of its UV continuum to be
$\alpha_{\rm \lambda}=-1.35\pm0.26$, which agrees but is slightly
flatter than the typical continuum slope of quasars. The estimated
rest-frame UV absolute magnitude of PSO J006+39 at the epoch of
its Keck/NIRSPEC observations is $M$$_{\rm 1450}=-25.60\pm0.07$.
It might differ at different epochs depending on the brightness
state of the quasar.

2. The mass of the SMBH of PSO J006+39 estimated using the CIV
line and the relation of \citetalias{2006ApJ...641..689V} is
$M_{\rm BH}=(2.19\pm0.30)\times10^{8}$$M_{\bigodot}$ and is about
3--4 times smaller than the typical masses of the SMBHs of
high-redshift quasars within a luminosity range of $-26.5<M_{\rm
1450}<-25.5$.

3. The SMBH of PSO J006+39 accretes at $\gtrsim$2 times of its
Eddington luminosity, which is higher than the typical accretion
rates of high-redshift quasars of similar luminosities. The high
accretion rate of the SMBH of PSO J006+39 suggests that it is
likely in the early stage of its formation, observed during the
phase of the intensive growth through accretion of surrounding
gas. As indicated by the negligible blueshift of the CIV emission
line of PSO J006+39, the quasar outflow in PSO J006+39 is probably
not very significant and does not prevent its SMBH from the
efficient and rapid growth. We might also suggest that star
formation in PSO J006+39 is probably less affected by the quasar
feedback than in other known luminous high-redshift quasars.

4. The metallicity of the circumnuclear gas of PSO J006+39 of
$\gtrsim$10$Z_{\bigodot}$ indicated by the NV/CIV line ratio is
somewhat high for the mass of its SMBH. Such high metallicity is
unlikely in the early phases of black hole formation. On the other
hand, the NV/CIV line ratio of PSO J006+39 might not be due to its
high metallicity. It might reflect a particular evolutionary phase
characterized by the high abundance of nitrogen. This high
abundance of nitrogen might have been produced by the
post-starburst population of stars which could trigger the quasar
activity of PSO J006+39 by providing the fuel for black hole
accretion. The Ly$\alpha$/CIV and CIII]/CIV line ratios also
indicate a lower level of ionization of the circumnuclear gas than
usually in quasars, which is more typical for NLSy1 galaxies, --
the low-redshift analogues of high-redshift quasars. Thus, the
evolutionary phase of PSO J006+39 differs from that of other known
high-redshift quasars of similar luminosities. Its intensively
growing black hole might be in the early phase of quasar activity.

\acknowledgments

The data presented herein were obtained at the W. M. Keck
Observatory (Program U055, PI: M. Malkan), which is operated as a
scientific partnership among the California Institute of
Technology, the University of California and the National
Aeronautics and Space Administration. The Observatory was made
possible by the generous financial support of the W. M. Keck
Foundation. Support for this work was provided by the Ministry of
Science and Technology of Taiwan, grant Nos MOST
105-2119-M-007-022-MY3, MOST 107-2119-M-008-009-MY3, and MOST
107-2811-M-008-2524. The authors wish to recognize and acknowledge
the very significant cultural role and reverence that the summit
of Maunakea has always had within the indigenous Hawaiian
community.  We are most fortunate to have the opportunity to
conduct observations from this mountain.


\end{document}